\documentclass[12pt]{article}
\usepackage{amssymb}
\font\oneeight=cmr10 at 18pt

\newcommand{\vTm}{\vphantom{\mbox{\oneeight I}}}

\begin{document}



\def\a{\alpha}
\def\b{\beta}
\def\d{\delta}
\def\e{\epsilon}
\def\g{\gamma}
\def\h{\mathfrak{h}}
\def\k{\kappa}
\def\l{\lambda}
\def\o{\omega}
\def\p{\wp}
\def\r{\rho}
\def\t{\tau}
\def\s{\sigma}
\def\z{\zeta}
\def\x{\xi}
 \def\A{{\cal{A}}}
 \def\B{{\cal{B}}}
 \def\C{{\cal{C}}}
 \def\D{{\cal{D}}}
\def\G{\Gamma}
\def\K{{\cal{K}}}
\def\O{\Omega}
\def\R{\bar{R}}
\def\T{{\cal{T}}}
\def\L{\Lambda}
\def\f{E_{\tau,\eta}(sl_2)}
\def\E{E_{\tau,\eta}(sl_n)}
\def\Zb{\mathbb{Z}}
\def\Cb{\mathbb{C}}

\def\R{\overline{R}}

\def\no{\nonumber}
\def\le{\langle}
\def\re{\rangle}
\def\lt{\left}
\def\rt{\right}

\baselineskip=20pt

\newfont{\elevenmib}{cmmib10 scaled\magstep1}
\newcommand{\preprint}{
   \begin{flushleft}
     \elevenmib Yukawa\, Institute\, Kyoto\\
   \end{flushleft}\vspace{-1.3cm}
   \begin{flushright}\normalsize
   \sf  YITP-04-46\\
     {\tt hep-th/0409002} \\ September 2004
   \end{flushright}}
\newcommand{\Title}[1]{{\baselineskip=26pt
   \begin{center} \Large \bf #1 \\ \ \\ \end{center}}}
\newcommand{\Author}{\begin{center}
   \large \bf
Wen-Li Yang$,{}^{a,b}$
 ~ Ryu Sasaki${}^c$ and~Yao-Zhong Zhang ${}^b$\end{center}}
\newcommand{\Address}{\begin{center}
     ${}^a$ Institute of Modern Physics, Northwest University,
     Xian 710069, P.R. China\\
     ${}^b$ Department of Mathematics, University of Queensland, Brisbane, QLD 4072,
     Australia\\
     ${}^c$ Yukawa Institute for Theoretical Physics, Kyoto
     University, Kyoto 606-8502, Japan
   \end{center}}
\newcommand{\Accepted}[1]{\begin{center}
   {\large \sf #1}\\ \vspace{1mm}{\small \sf Accepted for Publication}
   \end{center}}

\preprint
\thispagestyle{empty}
\bigskip\bigskip\bigskip

\Title{$\Zb_n$ elliptic  Gaudin
      model with open boundaries } \Author

\Address
\vspace{1cm}

\begin{abstract}
The $\Zb_n$ elliptic Gaudin model with  integrable boundaries
specified by generic {\it non-diagonal\/} K-matrices with $n+1$
free boundary parameters is studied. The commuting families of
Gaudin operators are diagonalized by the algebraic Bethe ansatz
method. The eigenvalues and the corresponding Bethe ansatz
equations are obtained.

\vspace{1truecm} \noindent {\it PACS:} 03.65.Fd; 04.20.Jb;
05.30.-d; 75.10.Jm

\end{abstract}
\newpage
\section{Introduction}
\label{intro} \setcounter{equation}{0}

In the study of one-dimensional many-body systems with long-range
interactions  Gaudin type models \cite{Gau76} stand out as a
particularly important class due to their applications in many
branches of physics such as the BCS theory \cite{Bar57} of small
metallic grains \cite{Cam97,Ami01, Del01}, theoretical nuclear
physics \cite{Dea03,Duk04}, quantum chromo-dynamics (QCD) theory
\cite{Iac94, Ris00} and Seiberg-Witten theory of supersymmetric
gauge theory  \cite{Sei94}. They also provide a powerful way for
the construction of integral representations of solutions to the
Knizhnik-Zamolodchikov (KZ) equation
\cite{Bab93,Has94,Fei94,Hik95,Gou02}.

Rational and trigonometric Gaudin magnets have been extensively
investigated in the literature. It is well-known that  the
periodic Gaudin's magnet Hamiltonians (or Gaudin operators) can be
constructed via the quasi-classical expansion of the transfer
matrix ({\it row-to-row transfer matrix\/}) of an inhomogeneous
spin chain with periodic boundary conditions \cite{Hik92,Skl96}.
Gaudin models with non-trivial boundary conditions can be in
principle  treated by means of the Sklyanin's boundary inverse
scattering method \cite{Skl88}. In this case the quasi-classical
expansion of the corresponding boundary transfer matrix ({\it
double-row transfer matrix\/}) produces {\it generalized\/} Gaudin
Hamiltonians with boundaries specified by certain K-matrices
\cite{Hik95, Lor02}. In particular, twisted boundary conditions
and open boundary conditions associated with {\it diagonal\/}
K-matrices give rise to Gaudin magnets in non-uniform local
magnetic fields \cite{Hik95} and interacting electron pairs with
certain non-uniform long-range coupling strengths
\cite{Ami01,Duk01,Zho02,Del02,Lor02}. Very recently, the XXZ
Gaudin model with generic integrable boundaries specified by
generic {\it non-diagonal\/} K-matrices \cite{Dev93, Gho94} was
solved by the algebraic Bethe ansatz method \cite{Yan041}.

So far, study of Gaudin type models via the {\em algebraic Bethe
ansatz method\/} is quite advanced for  rational and trigonometric
interaction cases. On the other hand, the elliptic interaction
case is rather less developed. We are aware of only two works
\cite{Skl96,Gou02} where the XYZ Gaudin model and its face-type
counterpart were respectively constructed and solved. In this
paper we will solve, using the algebraic Bethe ansatz method, the
most {\it generic\/} case--the $\Zb_n$ elliptic Gaudin magnets
with open boundary conditions specified by the generic {\it
non-diagonal\/} K-matrices given in \cite{Fan98,Yan03}, which
contain $n+1$ free boundary parameters. In section 3, we construct
the elliptic Gaudin operators associated with the generic boundary
K-matrices. The commutativity of these operators follows from the
standard procedure \cite{Hik92,Hik95,Lor02} specializing to the
inhomogeneous $\Zb_n$ Belavin  model with open boundaries
\cite{Bel81,Yan04}, thus ensuring the integrability of the $\Zb_n$
elliptic Gaudin magnets. In section 4, we diagonalize the Gaudin
operators simultaneously by means of the algebraic Bethe ansatz
method. This constitutes the main new result in this paper. The
diagonalization is achieved by means of the technique of the
``vertex-face" transformation. In section 5, we conclude this
paper with a summary and comments.


\section{ Preliminaries: the inhomogeneous $\Zb_n$ Belavin model
with open boundaries}
\label{Zn} \setcounter{equation}{0}

Let us fix a positive integer $n\geq 2$, a complex number $\tau$
such that $Im(\tau)>0$ and a generic complex number $w$. Introduce
the following elliptic functions
\begin{eqnarray}
\theta\lt[
 \begin{array}{c}
 a\\b
 \end{array}\rt](u,\tau)&=&\sum_{m=-\infty}^{\infty}
 exp\lt\{\sqrt{-1}\pi\lt[(m+a)^2\tau+2(m+a)(u+b)\rt]\rt\},\\
 \theta^{(j)}(u)&=&\theta\lt[\begin{array}{c}\frac{1}{2}-\frac{j}{n}\\
 [2pt]\frac{1}{2}
 \end{array}\rt](u,n\tau),\qquad
 \s(u)=\theta\lt[\begin{array}{c}\frac{1}{2}\\[2pt]\frac{1}{2}
 \end{array}\rt](u,\tau),\label{Function}\\
 \zeta(u)&=&\frac{\partial}{\partial u}\lt\{\ln
 \s(u)\rt\}.\label{Z-function}
\end{eqnarray}
Among them the
$\s$-function\footnote{Our $\s$-function is the
$\vartheta$-function $\vartheta_1(u)$ \cite{Whi50}. It has the
following relation with the {\it Weierstrassian\/} $\s$-function
if denoted it by
$\s_w(u)$: $\s_w(u)\propto e^{\eta_1u^2}\s(u)$,
$\eta_1=\pi^2(\frac{1}{6}-4\sum_{n=1}^{\infty}\frac{nq^{2n}}{1-q^{2n}})
$ and $q=e^{\sqrt{-1}\tau}$. Consequently, our $\zeta$-function
given by (\ref{Z-function}) is different from the {\it
Weierstrassian\/} $\zeta$-function by an additional term $-2\eta_1
u$.}
 satisfies the following
identity:
\begin{eqnarray}
 &&\s(u+x)\s(u-x)\s(v+y)\s(v-y)-\s(u+y)\s(u-y)\s(v+x)\s(v-x)\no\\
 &&\qquad\qquad \qquad =\s(u+v)\s(u-v)\s(x+y)\s(x-y).\no
\end{eqnarray}
Let $g,\,h,$ be
$n\times n$ matrices with the elements
\begin{eqnarray}
 &&h_{ij}=\d_{i+1\,j},~~g_{ij}=\o^i\d_{i\,j},~~{\rm
 with}\quad \o=e^{\frac{2\pi\sqrt{-1}}{n}},~~~i,j\in \Zb_n.\no
\end{eqnarray}
For any $\a=(\a_1,\a_2)$, $\a_1,\,\a_2\in \Zb_n$, one can introduce an
associated $n\times n$ matrix $I_{\a}$ defined by
\begin{eqnarray}
 &&I_{\a}=I_{(\a_1,\a_2)}=g^{\a_2}h^{\a_1},\no
\end{eqnarray}
and an elliptic function $\s_{\a}(u)$ given by
\begin{eqnarray}
 &&\s_{\a}(u)=\theta\lt[\begin{array}{c}\frac{1}{2}+\frac{\a_1}{n}\\[2pt]
 \frac{1}{2}+\frac{\a_2}{n}
 \end{array}\rt](u,\tau),~~{\rm and\/}~\s_{(0,0)}(u)=\s(u).\no
\end{eqnarray}
The $\Zb_n$ Belavin R-matrix  is given by \cite{Bel81}
\begin{eqnarray}
 R^B(u)=\frac{\s(w)}{\s(u+w)}\sum_{\a\in\Zb_n^2}
 \frac{\s_{\a}(u+\frac{w}{n})}
 {n\s_{\a}(\frac{w}{n})}I_{\a}\otimes
 I_{\a}^{-1}.\label{Belavin-R}
\end{eqnarray}
The R-matrix satisfies the
quantum Yang-Baxter equation (QYBE)
\begin{eqnarray}
 R_{12}(u_1-u_2)R_{13}(u_1-u_3)R_{23}(u_2-u_3)=
 R_{23}(u_2-u_3)R_{13}(u_1-u_3)R_{12}(u_1-u_2), \label{QYB}
\end{eqnarray}
and the properties \cite{Ric86},
\begin{eqnarray}
 &&\hspace{-1.5cm}\mbox{
 Unitarity}:\hspace{42.5mm}R^B_{12}(u)R^B_{21}(-u)= {\rm id},\label{Unitarity}\\
 &&\hspace{-1.5cm}\mbox{
 Crossing-unitarity}:\quad (R^B)^{t_2}_{21}(-u-nw)(R^B)_{12}^{t_2}(u)
 = \frac{\s(u)\s(u+nw)}{\s(u+w)\s(u+nw-w)}\,\mbox{id},
 \label{crosing-unitarity}\\
 &&\hspace{-1.5cm}\mbox{ Quasi-classical
 property}:\hspace{22.5mm}\, R^B_{12}(u)|_{w\rightarrow 0}= {\rm
id}.\label{quasi}
\end{eqnarray}
Here $R^B_{21}(u)=P_{12}R^B_{12}(u)P_{12}$ with $P_{12}$
being the usual permutation operator and $t_i$ denotes the
transposition in the $i$-th space. Here and below we adopt the
standard notation: for any matrix $A\in {\rm End}(\Cb^n)$, $A_j$
is an embedding operator in the tensor space $\Cb^n\otimes
\Cb^n\otimes\cdots$, which acts as $A$ on the $j$-th space and as
an identity on the other factor spaces; $R_{ij}(u)$ is an
embedding operator of R-matrix in the tensor space, which acts as
an identity on the factor spaces except for the $i$-th and $j$-th
ones. The quasi-classical properties (\ref{quasi}) of the R-matrix
enables one to introduce an associated $\Zb_n$ classical r-matrix
$r(u)$ \cite{Hou99}
\begin{eqnarray}
 R^B(u)&=&{\rm id} +w\,r(u)+O(w^2),\qquad\qquad  {\rm
 when}~w\longrightarrow 0,\no\\[4pt]
 r(u)&=&\frac{1-n}{n}\zeta(u)+\hspace{-0.2truecm}
 \sum_{\a\in\Zb_n^2-(0,0)}\frac{\s'(0)\s_{\a}(u)}
 {n\s(u)\s_{\a}(0)}I_{\a}\otimes
 I_{\a}^{-1},~~\s'(0)=\frac{\partial}{\partial
 u}\s(u)|_{u=0}.\label{r-matrix}
\end{eqnarray}
In the above equation, the
elliptic $\zeta$-function is defined in (\ref{Z-function}).

One introduces  the ``row-to-row" monodromy matrix $T(u)$
\cite{Kor93}, which is an $n\times n$ matrix with elements being
operators acting  on $(\Cb^n)^{\otimes N}$  \begin{eqnarray}
T(u)=R^B_{01}(u+z_1)R^B_{02}(u+z_2)\cdots
R^B_{0N}(u+z_N).\label{T-matrix}\end{eqnarray} Here $\{z_i|i=1,\cdots, N\}$
are arbitrary free complex parameters which are usually called
inhomogeneous parameters. With the help of the QYBE (\ref{QYB}),
one can show that $T(u)$ satisfies the so-called ``RLL" relation
\begin{eqnarray}
R^B_{12}(u-v)T_1(u)T_2(v)=T_2(v)T_1(u)R^B_{12}(u-v).\label{Relation1}\end{eqnarray}

An integrable open chain can be constructed as follows
\cite{Skl88}. Let us introduce a pair of K-matrices $K^-(u)$ and
$K^+(u)$. The former satisfies the reflection equation (RE)
 \begin{eqnarray}
 &&R^B_{12}(u_1-u_2)K^-_1(u_1)R^B_{21}(u_1+u_2)K^-_2(u_2)\no\\
  &&~~~~~~=
 K^-_2(u_2)R^B_{12}(u_1+u_2)K^-_1(u_1)R^B_{21}(u_1-u_2),\label{RE-V}
\end{eqnarray}
and the latter  satisfies the dual RE
\begin{eqnarray}
 &&R^B_{12}(u_2-u_1)K^+_1(u_1)R^B_{21}(-u_1-u_2-nw)K^+_2(u_2)\no\\
 &&~~~~~~=
 K^+_2(u_2)R^B_{12}(-u_1-u_2-nw)K^+_1(u_1)R^B_{21}(u_2-u_1).
 \label{DRE-V}
\end{eqnarray}
For the models with open boundaries, instead of
the standard ``row-to-row" monodromy matrix $T(u)$
(\ref{T-matrix}), one needs  the
 ``double-row" monodromy matrix $\mathbb{T}(u)$
\begin{eqnarray}
 \mathbb{T}(u)=T(u)K^-(u)T^{-1}(-u).\label{Mon-V-1}
\end{eqnarray}
Using
(\ref{Relation1}) and (\ref{RE-V}), one can prove that
$\mathbb{T}(u)$ satisfies
\begin{eqnarray}
 R^B_{12}(u_1-u_2)\mathbb{T}_1(u_1)R^B_{21}(u_1+u_2)
  \mathbb{T}_2(u_2)=
 \mathbb{T}_2(u_2)R^B_{12}(u_1+u_2)\mathbb{T}_1(u_1)R^B_{21}(u_1-u_2).
 \label{Relation-Re}
\end{eqnarray}
Then the {\it double-row transfer
matrix\/} of  the inhomogeneous $\Zb_n$ Belavin model  with open
boundary is given by
\begin{eqnarray}
\t(u)=tr(K^+(u)\mathbb{T}(u)).\label{trans}
\end{eqnarray}
The commutativity of the transfer matrices
\begin{eqnarray}
 [\t(u),\t(v)]=0,\label{Com-2}
\end{eqnarray}
follows as a consequence of (\ref{QYB})-(\ref{crosing-unitarity})
and (\ref{RE-V})-(\ref{DRE-V}). This ensures the integrability of
the inhomogeneous $\Zb_n$ Belavin model with open boundary.
\section{$\Zb_n$ elliptic  Gaudin models with generic boundaries}
 \label{Haml} \setcounter{equation}{0}
In this paper, we will consider a {\it non-diagonal\/} K-matrix
$K^{-}(u)$ which is a solution to the RE (\ref{RE-V}) associated
with the $\Zb_n$ Belavin R-matrix \cite{Fan98}
\begin{eqnarray}
 K^-(u)^s_t=\sum_{i=1}^n \frac{\s(\l_i+\xi-u)}{\s(\l_i+\xi+u)}
 \,\phi^{(s)}_{\l,\l-w\hat{\imath}}(u)
 \,\bar{\phi}^{(t)}_{\l,\l-w\hat{\imath}}(-u). \label{K-matrix}
\end{eqnarray}
The corresponding {\it dual\/} K-matrix $K^+(u)$ which is a
solution to the dual RE (\ref{DRE-V}) has been obtained in
\cite{Yan03}.  With a particular choice of the free boundary
parameters with respect to $K^-(u)$, we have
\begin{eqnarray}
 K^+(u)^s_t&=&
 \sum_{i=1}^n \lt\{\prod_{k\neq
 i}\frac{\s((\l_i-\l_k)-w)}{\s(\l_i-\l_k)}\rt\}
 \frac{\s(\l_i+\bar{\xi}+u+
 \frac{nw}{2})}{\s(\l_i+\bar{\xi}-u-\frac{nw}{2})}\no\\[2pt]
 &&\qquad \times \phi^{(s)}_{\l,\l-w\hat{\imath}}(-u)\,
 \tilde{\phi}^{(t)}_{\l,\l-w\hat{\imath}}(u).\label{K-matrix1}
\end{eqnarray}
In (\ref{K-matrix}) and (\ref{K-matrix1}),
$\phi,\,\bar{\phi},\,\tilde{\phi}$ are intertwiners which will be
specified  in section 4. We consider the generic $\{\l_i\}$ such
that $\l_i\neq \l_j\,\, (modulo ~\Zb+\tau\Zb)$ for $i\neq j$. This
condition is necessary for the non-singularity of $K^{\pm}(u)$. It
is convenient to introduce a vector $\l=\sum_{i=1}^n\l_i\e_i$
associated with the boundary parameters $\{\l_i\}$, where
$\{\e_i,~i=1,\cdots,n\}$ is the orthonormal basis of the vector
space $\Cb^n$ such that $\langle \e_i,\e_j\rangle=\d_{ij}$.

 As will be seen from the definitions of the intertwiners
(\ref{Intvect}), (\ref{Int1}) and (\ref{Int2}) specialized to
$m=\l$, $\phi_{\l,\l-w\hat{\imath}}(u)$ and
$\bar{\phi}_{\l,\l-w\hat{\imath}}(u)$ do not depend on $w$ but
$\tilde{\phi}_{\l,\l-w\hat{\imath}}(u)$ does. Consequently, the
K-matrix $K^-(u)$ does not depend on the crossing parameter $w$,
 but $K^+(u)$ does. So we use the
convention:
\begin{eqnarray}
 K(u)=\lim_{w\rightarrow
 0}K^-(u)=K^-(u).\label{Conv1}
\end{eqnarray}
We further restrict the complex
parameters $\xi$ and $\bar{\xi}$ to be the same, i.e.\begin{eqnarray}
\bar{\xi}=\xi,\label{Restriction-1}\end{eqnarray} so that (\ref{ID-1}) below
is satisfied. Hence, the K-matrices depend on $n+1$ free
parameters $\{\l_i|i=1,\cdots,n\}$ and $\xi$, which specify the
integrable boundary conditions \cite{Gho94}. Moreover, the
K-matrices satisfy the following relation thanks to  the
restriction (\ref{Restriction-1})
\begin{eqnarray}
\lim_{w\rightarrow
0}\{K^+(u)\,K^-(u)\}=\lim_{w\rightarrow 0}\{K^+(u)\}K(u)={\rm
id}.\label{ID-1}
\end{eqnarray}

Let us introduce the $\Zb_n$ elliptic  Gaudin operators $\{H_j
|j=1,2,\cdots,N\}$ associated with the inhomogeneous $\Zb_n$
Belavin  model with open boundaries specified by the generic
K-matrices (\ref{K-matrix}) and (\ref{K-matrix1}):
\begin{eqnarray}
H_j=\G_j(z_j)+\sum_{k\neq
 j}^{N}r_{kj}(z_j-z_k)+K^{-1}_j(z_j)\lt\{\sum_{k\neq
 j}^{N}r_{jk}(z_j+z_k)\rt\}K_j(z_j),\label{Ham}
\end{eqnarray}
where
$\G_j(u)=\frac{\partial}{\partial
w}\{\bar{K}_j(u)\}|_{w=0}K_j(u)$, $j=1,\cdots,N,$ with
$\bar{K}_j(u)=tr_0\lt\{K^+_0(u)R^B_{0j}(2u)P_{0j}\rt\}$. Here
$\{z_j\}$ are the inhomogeneous parameters of the inhomogeneous
$\Zb_n$ Belavin model and $r(u)$ is given by (\ref{r-matrix}). For
a generic choice of the boundary parameters
$\{\l_1,\,\cdots,\l_n,\,\xi\}$, $\G_j(u)$ is a {\it
non-diagonal\/} matrix.

The elliptic  Gaudin operators (\ref{Ham}) are obtained by
expanding the double-row transfer matrix (\ref{trans}) at the
point $u=z_j$ around $w=0$:
\begin{eqnarray}
 \t(z_j)&=&\t(z_j)|_{w=0}+w
 H_j+O(w^2),\quad j=1,\cdots,N, \label{trans-2}\\
 H_j&=&\frac{\partial}{\partial w}\t(z_j)|_{w=0}\,  .\label{Eq-1}
\end{eqnarray}
The relations (\ref{quasi}) and (\ref{ID-1}) imply that the first
term $\t(z_j)|_{w=0}$ in the expansion (\ref{trans-2}) is equal to
an identity, namely,
\begin{eqnarray}
 \t(z_j)|_{w=0}={\rm id}.\label{First}
\end{eqnarray}
Then the commutativity of the transfer matrices $\{\t(z_j)\}$
(\ref{Com-2}) for a generic $w$ implies
\begin{eqnarray}
[H_j,H_k]=0,\quad i,j=1,\cdots,N.\label{Com-1}
\end{eqnarray} Thus the elliptic
Gaudin system defined by (\ref{Ham}) is integrable. Moreover, the
fact that the Gaudin operators $\{H_j\}$ (\ref{Ham}) can be
expressed in terms of the transfer matrix of the inhomogeneous
$\Zb_n$ Belavin model with open boundary enables us to exactly
diagonalize the operators by the algebraic Bethe ansatz method
with the help of the ``vertex-face" correspondence technique, as
will be shown in the next section. The aim of this paper is to
diagonalize  the elliptic Gaudin operators $H_j$, $j=1,\cdots,N$
(\ref{Ham}) simultaneously.


\section{Eigenvalues and Bethe ansatz equations}
\label{BAE} \setcounter{equation}{0}
\subsection{$A^{(1)}_{n-1}$ SOS R-matrix and face-vertex
correspondence}

The $A_{n-1}$ simple roots $\{\a_i\}$ can be expressed in terms of
the orthonormal basis $\{\e_i\}$  as:
\begin{eqnarray}
 \a_{i}=\e_i-\e_{i+1},\quad i=1,\cdots,n-1,\no
\end{eqnarray}
and the fundamental
weights $\lt\{\L_i~|~i=1,\cdots,n-1\rt\}$ satisfying
$\langle\L_i,~\a_j\rangle=\d_{ij}$ are given by
\begin{eqnarray}
 \L_i=\sum_{k=1}^{i}\e_k-\frac{i}{n}\sum_{k=1}^{n}\e_k. \no
\end{eqnarray}
Set
\begin{eqnarray}
 \hat{\imath}=\e_i-\overline{\e},\quad \overline{\e}=
 \frac{1}{n}\sum_{k=1}^{n}\e_k,\quad i=1,\cdots,n,\quad {\rm
 then}\quad \sum_{i=1}^n\hat{\imath}=0. \label{Vectors}
\end{eqnarray}
For each
dominant weight $\L=\sum_{i=1}^{n-1}a_i\L_{i}$,\ $a_{i}\in
\Zb^+$ (the set of non-negative integers), there exists an
irreducible highest weight finite-dimensional representation
$V_{\L}$ of $A_{n-1}$ with the highest vector $ |\L\rangle$. For
example the fundamental vector representation is $V_{\L_1}$.

Let $\h$ be the Cartan subalgebra of $A_{n-1}$ and $\h^{*}$ be its
dual. A finite dimensional diagonalizable  $\h$-module is a
complex finite dimensional vector space $W$ with a weight
decomposition $W=\oplus_{\mu\in \h^*}W[\mu]$, so that $\h$ acts on
$W[\mu]$ by $x\,v=\mu(x)\,v$, $(x\in \h,\,v\in\,W[\mu])$. For
example, the fundamental vector representation $V_{\L_1}=\Cb^n$,
the non-zero weight spaces $W[\hat{\imath}]=\Cb
\e_i,~i=1,\cdots,n$.

For a generic $m\in \Cb^n$, define
\begin{eqnarray}
 m_i=\langle m,\e_i\rangle,
 \quad m_{ij}=m_i-m_j=\langle m,\e_i-\e_j\rangle,\quad i,j=1,\cdots,n.
 \label{Def1}
\end{eqnarray}
Let $R(u,m)\in End(\Cb^n\otimes\Cb^n)$ be the
R-matrix of the $A^{(1)}_{n-1}$ SOS model \cite{Jim87} given by
\begin{eqnarray}
R(u,m)=\sum_{i=1}^{n}
 R^{ii}_{ii}(u,m)E_{ii}
\otimes
 E_{ii} +
 \sum_{i\ne j}
\lt\{R^{ij}_{ij}(u,m)E_{ii}
\otimes
E_{jj}+
 R^{ji}_{ij}(u,m)E_{ji}\otimes
 E_{ij}\rt\}, \label{R-matrix}
\end{eqnarray}
where $E_{ij}$ is the matrix
with elements $(E_{ij})^l_k=\d_{jk}\d_{il}$. The coefficient
functions are
\begin{eqnarray}
 &&R^{ii}_{ii}(u,m)=1,\qquad
 R^{ij}_{ij}(u,m)=\frac{\s(u)\s(m_{ij}-w)}
 {\s(u+w)\s(m_{ij})},\quad i\neq j,\label{Elements1}\\
 && R^{j\,i}_{ij}(u,m)=\frac{\s(w)\s(u+m_{ij})}
 {\s(u+w)\s(m_{ij})},\quad i\neq j,\label{Elements2}
\end{eqnarray}
and  $m_{ij}$
are defined in (\ref{Def1}). The R-matrix satisfies the dynamical
(modified) QYBE
\begin{eqnarray}
&&R_{12}(u_1-u_2,m-wh^{(3)})R_{13}(u_1-u_3,m)
R_{23}(u_2-u_3,m-wh^{(1)})\no\\
&&\qquad =R_{23}(u_2-u_3,m)R_{13}(u_1-u_3,m-wh^{(2)})R_{12}(u_1-u_2,m),
\label{MYBE}
\end{eqnarray}
and the quasi-classical property \begin{eqnarray}
R(u,m)|_{w\rightarrow 0}={\rm id}.\label{quasi1}\end{eqnarray} We adopt the
notation: $R_{12}(u,m-wh^{(3)})$ acts on a tensor $v_1\otimes v_2
\otimes v_3$ as $R(u,m-w\mu)\otimes id$ if $v_3\in W[\mu]$.
Moreover, the R-matrix satisfies the unitarity and the modified
crossing-unitarity relation \cite{Jim87}.

Let us introduce an intertwiner---an $n$-component  column vector
$\phi_{m,m-w\hat{\jmath}}(u)$ whose  $k$-th element is \begin{eqnarray}
\phi^{(k)}_{m,m-w\hat{\jmath}}(u)=\theta^{(k)}(u+nm_j).\label{Intvect}\end{eqnarray}
Using the intertwining vector, one  derives the following
face-vertex correspondence relation \cite{Jim87}
\begin{eqnarray}
 &&
 R^B_{12}(u_1-u_2)\, \phi_{m,m-w\hat{\imath}}(u_1)\otimes
 \phi_{m-w\hat{\imath},m-w(\hat{\imath}+\hat{\jmath})}(u_2)\no\\
 &&\qquad = \sum_{k,l}R(u_1-u_2,m)^{kl}_{ij}\,
 \phi_{m-w\hat{l},m-w(\hat{l}+\hat{k})}(u_1)\otimes
 \phi_{m,m-w\hat{l}}(u_2). \label{Face-vertex}
\end{eqnarray}
Then the QYBE of
$\Zb_n$ Belavin's R-matrix $R^B(u)$ (\ref{QYB}) is equivalent to
the dynamical Yang-Baxter equation of $A^{(1)}_{n-1}$ SOS R-matrix
$R(u,m)$ (\ref{MYBE}). For a generic $m$, we may introduce other
types of intertwiners $\bar{\phi},~\tilde{\phi}$ satisfying the
conditions,
\begin{eqnarray}
 &&\sum_{k=1}^n\bar{\phi}^{(k)}_{m,m-w\hat{\mu}}(u)
 ~\phi^{(k)}_{m,m-w\hat{\nu}}(u)=\d_{\mu\nu},\label{Int1}\\
 &&\sum_{k=1}^n\tilde{\phi}^{(k)}_{m+w\hat{\mu},m}(u)
 ~\phi^{(k)}_{m+w\hat{\nu},m}(u)=\d_{\mu\nu},\label{Int2}
\end{eqnarray}
{}from which one  derives the relations, \begin{eqnarray}
&&\sum_{\mu=1}^n\bar{\phi}^{(i)}_{m,m-w\hat{\mu}}(u)
~\phi^{(j)}_{m,m-w\hat{\mu}}(u)=\d_{ij},\label{Int3}\\
&&\sum_{\mu=1}^n\tilde{\phi}^{(i)}_{m+w\hat{\mu},m}(u)
~\phi^{(j)}_{m+w\hat{\mu},m}(u)=\d_{ij}.\label{Int4}\end{eqnarray} With the
help of (\ref{Int1})-(\ref{Int4}), we obtain the following
relations from the face-vertex correspondence relation
(\ref{Face-vertex}):
\begin{eqnarray}
 &&\left(\tilde{\phi}_{m+w\hat{k},m}(u_1)\otimes
 {\rm id}\right)R^B_{12}(u_1-u_2) \left(\vTm{\rm
id}\otimes\phi_{m+w\hat{\jmath},m}(u_2)\right)\no\\
 &&\qquad\quad= \sum_{i,l}R(u_1-u_2,m)^{kl}_{ij}\,
 \tilde{\phi}_{m+w(\hat{\imath}+\hat{\jmath}),m+w\hat{\jmath}}(u_1)\otimes
 \phi_{m+w(\hat{k}+\hat{l}),m+w\hat{k}}(u_2),\label{Face-vertex1}\\
 &&\left(\tilde{\phi}_{m+w\hat{k},m}(u_1)\otimes
 \tilde{\phi}_{m+w(\hat{k}+\hat{l}),m+w\hat{k}}(u_2)\right)R^B_{12}(u_1-u_2)\no\\
 &&\qquad\quad= \sum_{i,j}R(u_1-u_2,m)^{kl}_{ij}\,
 \tilde{\phi}_{m+w(\hat{\imath}+\hat{\jmath}),m+w\hat{\jmath}}(u_1)\otimes
 \tilde{\phi}_{m+w\hat{\jmath},m}(u_2),\label{Face-vertex2}\\
 &&\left({\rm id}\otimes
 \bar{\phi}_{m,m-w\hat{l}}(u_2)\right)R^B_{12}(u_1-u_2)
 \left(\vTm\phi_{m,m-w\hat{\imath}}(u_1)\otimes {\rm id}\right)\no\\
 &&\qquad\quad= \sum_{k,j}R(u_1-u_2,m)^{kl}_{ij}\,
 \phi_{m-w\hat{l},m-w(\hat{k}+\hat{l})}(u_1)\otimes
 \bar{\phi}_{m-w\hat{\imath},m-w(\hat{\imath}+\hat{\jmath})}(u_2),\label{Face-vertex3}\\
 &&\left(\bar{\phi}_{m-w\hat{l},m-w(\hat{k}+\hat{l})}(u_1)\otimes
 \bar{\phi}_{m,m-w\hat{l}}(u_2)\right)R^B_{12}(u_1-u_2)\no\\
 &&\qquad\quad= \sum_{i,j}R(u_1-u_2,m)^{kl}_{ij}\,
 \bar{\phi}_{m,m-w\hat{\imath}}(u_1)\otimes
 \bar{\phi}_{m-w\hat{\imath},m-w(\hat{\imath}
 +\hat{\jmath})}(u_2).\label{Face-vertex4}
\end{eqnarray}
The face-vertex correspondence relations (\ref{Face-vertex})
and (\ref{Face-vertex1})-(\ref{Face-vertex4}) will play an
important role in {\it translating\/}  formulas in the ``vertex
form" into those in the ``face form" so that the algebraic Bethe
ansatz method can be applied to diagonalize the transfer matrix.

Corresponding to the vertex type K-matrices (\ref{K-matrix}) and
(\ref{K-matrix1}), one introduces the following face type
K-matrices $\K$ and $\tilde{\K}$, as in \cite{Yan03}
\begin{eqnarray}
 &&\K(\l|u)^j_i=\sum_{s,t}\tilde{\phi}^{(s)}
 _{\l-w(\hat{\imath}-\hat{\jmath}),~\l-w\hat{\imath}}(u)\,K(u)^s_t\,\phi^{(t)}
 _{\l,~\l-w\hat{\imath}}(-u),\label{K-F-1}\\
 &&\tilde{\K}(\l|u)^j_i=\sum_{s,t}\bar{\phi}^{(s)}
 _{\l,~\l-w\hat{\jmath}}(-u)\,\tilde{K}(u)^s_t\,\phi^{(t)}
 _{\l-w(\hat{\jmath}-\hat{\imath}),~\l-w\hat{\jmath}}(u).\label{K-F-2}
\end{eqnarray}
Through straightforward calculations, we find the face type
K-matrices {\it simultaneously\/} have  {\it diagonal\/}
forms\footnote{As will be seen below (\ref{Parameters}), the
spectral parameter $u$ and the boundary parameter $\xi$ of the
reduced double-row monodromy matrices constructed from $\K(\l|u)$
will be shifted in each step of the nested Bethe ansatz procedure
\cite{Yan04}. Therefore, it is convenient to specify the
dependence on the boundary parameter $\xi$ of $\K(\l|u)$ in
addition to the spectral parameter $u$. }
\begin{eqnarray}
\K(\l|u)^j_i=\d_i^j\,k(\l|u;\xi)_i,\quad
\tilde{\K}(\l|u)^j_i=\d_i^j\,\tilde{k}(\l|u)_i,\label{Diag-F}
\end{eqnarray}
where functions $k(\l|u;\xi)_i,\,\tilde{k}(\l|u)_i $ are given by
\begin{eqnarray}
 k(\l|u;\xi)_i
 &=&\frac{\s(\l_i+\xi-u)}{\s(\l_i+\xi+u)},\label{k-def}\\
 \tilde{k}(\l|u)_i&=&\lt\{\prod_{k\neq
 i,k=1}^n\frac{\s(\l_{ik}-w)}{\s(\l_{ik})}\rt\}
 \frac{\s(\l_i+\bar{\xi}+u+
 \frac{nw}{2})}{\s(\l_i+\bar{\xi}-u-\frac{nw}{2})}.
 \label{k-def1}
\end{eqnarray}
Moreover, one can check that the matrices
$\K(\l|u)$ and $\tilde{\K}(\l|u)$ satisfy the SOS type reflection
equation and its dual, respectively \cite{Yan03}. Although the
K-matrices $K^{\pm}(u)$ given by (\ref{K-matrix}) and
(\ref{K-matrix1}) are generally non-diagonal (in the vertex form),
after the face-vertex transformations (\ref{K-F-1}) and
(\ref{K-F-2}), the face type counterparts $\K(\l|u)$ and
$\tilde{\K}(\l|u)$ {\it simultaneously\/} become diagonal. This
fact enables the authors  to apply the generalized algebraic Bethe
ansatz method developed in \cite{Yan04} for SOS type integrable
models to diagonalize the transfer matrix $\t(u)$ (\ref{trans}).

\subsection{Algebraic Bethe ansatz}

By means of (\ref{Int3}), (\ref{Int4}), (\ref{K-F-2}) and
(\ref{Diag-F}), the transfer matrix $\t(u)$ (\ref{trans}) can be
recast  into the following face type form:
\begin{eqnarray}
 \t(u)&=&tr(K^+(u)\mathbb{T}(u))\no\\
 &=&\sum_{\mu,\nu}tr\lt(\!K^+(u)\, \phi_{\l-w(\hat{\mu}-\hat{\nu}),
 \l-w\hat{\mu}}(u)\,\tilde{\phi}_{\l-w(\hat{\mu}-\hat{\nu}),
 \l-w\hat{\mu}}(u)~\mathbb{T}(u) \phi_{\l, \l-w\hat{\mu}}(-u)
 \bar{\phi}_{\l, \l-w\hat{\mu}}(-u)\rt)\no\\
 &=&\sum_{\mu,\nu}\bar{\phi}_{\l, \l-w\hat{\mu}}(-u)K^+(u)
 \,\phi_{\l-w(\hat{\mu}-\hat{\nu}),\l-w\hat{\mu}}(u)~
 \tilde{\phi}_{\l-w(\hat{\mu}-\hat{\nu}),
 \l-w\hat{\mu}}(u)~\mathbb{T}(u)\phi_{\l, \l-w\hat{\mu}}(-u)\no\\
 &=&\sum_{\mu,\nu}\tilde{\K}(\l|u)_{\nu}^{\mu}\,\T(\l|u)^{\nu}_{\mu}=
 \sum_{\mu}\tilde{k}(\l|u)_{\mu}\,\T(\l|u)^{\mu}_{\mu}.
 \label{De1}
\end{eqnarray}
Here we have introduced the face-type double-row
monodromy matrix $\T(\l|u)$,
\begin{eqnarray}
 \T(\l|u)^{\nu}_{\mu}&=&\tilde{\phi}_{\l-w(\hat{\mu}-\hat{\nu}),
 \l-w\hat{\mu}}(u)~\mathbb{T}(u)\,\phi_{\l,
 \l-w\hat{\mu}}(-u)\no\\
 &\equiv&
 \sum_{i,j}\tilde{\phi}^{(j)}_{\l-w(\hat{\mu}-\hat{\nu}),
 \l-w\hat{\mu}}(u)~\mathbb{T}(u)^j_i\,\phi^{(i)}_{\l,
 \l-w\hat{\mu}}(-u).\label{Mon-F}
\end{eqnarray}
This face-type double-row
monodromy matrix can  be expressed in terms of the face type
R-matrix $R(u,\l)$ (\ref{R-matrix}) and the K-matrix $\K(\l|u)$
(\ref{K-F-1}) \cite{Yan04}. Moreover from (\ref{Relation-Re}),
(\ref{Face-vertex}) and (\ref{Int4}) one may derive the following
exchange relations among $\T(\l|u)^{\nu}_{\mu}$:
\begin{eqnarray}
 &&\sum_{i_1,i_2}\sum_{j_1,j_2}~
 R(u_1-u_2,\l)^{i_0,j_0}_{i_1,j_1}\,\T(\l+w(\hat{\jmath}_1+\hat{\imath}_2)|u_1)
 ^{i_1}_{i_2}\no\\
 &&~~~~~~~~\times R(u_1+u_2,\l)^{j_1,i_2}_{j_2,i_3}\,
 \T(\l+w(\hat{\jmath}_3+\hat{\imath}_3)|u_2)^{j_2}_{j_3}\no\\[2pt]
 &&~~=\sum_{i_1,i_2}\sum_{j_1,j_2}~
 \T(\l+w(\hat{\jmath}_1+\hat{\imath}_0)|u_2)
 ^{j_0}_{j_1}\,R(u_1+u_2,\l)^{i_0,j_1}_{i_1,j_2}\no\\
 &&~~~~~~~~\times\T(\l+w(\hat{\jmath}_2+\hat{\imath}_2)|u_1)^{i_1}_{i_2}\,
 R(u_1-u_2,\l)^{j_2,i_2}_{j_3,i_3}.\label{RE-F}
\end{eqnarray}

As in \cite{Yan04}, we introduce a standard notation for
convenience%
\footnote{The scalar factors in the definitions of
the operators $\B(\l|u)$ and $\D(\l|u)$ are to make the relevant
commutation relations as concise  as (\ref{Rel-1})-(\ref{Rel-3}).
}:
\begin{eqnarray}
\A(\l|u)&=&\T(\l|u)^1_1,\qquad \B_i(\l|u)=
 \frac{\T(\l|u)^1_i}{\s(\l_{i1})},\quad i=2,\cdots,n,\label{Def-AB}\\
\D^j_i(\l|u)&=&\frac{\s(\l_{j1}-\d_{ij}w)}{\s(\l_{i1})}
 \left\{\T(\l|u)^j_i-\d^j_iR(2u,\l+w\hat{1})^{j\,1}_{1\,j}\,\A(\l|u)\right\}
 ,\no\\
 &&\qquad i,j=2,\cdots,n. \label{Def-D}
\end{eqnarray}
{}From (\ref{RE-F}) one
may derive  the commutation relations among $\A(\l|u)$, $\D(\l|u)$
and $\B(\l|u)$. Here we give those which are relevant for our
purpose
\begin{eqnarray}
&&\A(\l|u)\B_i(\l+w(\hat{\imath}-\hat{1})|v)\no\\
&&\qquad=\frac{\s(u+v)\s(u-v-w)}{\s(u+v+w)\s(u-v)}
\,\B_i(\l+w(\hat{\imath}-\hat{1})|v)\A(\l+w(\hat{\imath}-\hat{1})|u)\no\\
&&\qquad\quad-\frac{\s(w)\s(2v)}{\s(u-v)\s(2v+w)}
\frac{\s(u-v-\l_{1i}+w)} {\s(\l_{1i}-w)}
\,\B_i(\l+w(\hat{\imath}-\hat{1})|u)\A(\l+w(\hat{\imath}-\hat{1})|v)\no\\
&&\qquad\quad-\frac{\s(w)}{\s(u+v+w)}\sum_{\a=2}^{n}
\frac{\s(u+v+\l_{\a1}+2w)}{\s(\l_{\a1}+w)}
\,\B_{\a}(\l+w(\hat{\a}-\hat{1})|u)\D^{\a}_i(\l+w(\hat{\imath}-\hat{1})|v),\no\\
\label{Rel-1}\\
&&\D^k_i(\l|u)\B_j(\l+w(\hat{\jmath}-\hat{1})|v)\no\\
&&\qquad=
\frac{\s(u-v+w)\s(u+v+2w)}{\s(u-v)\s(u+v+w)}\no\\
&&\qquad\qquad\qquad\qquad\qquad\times\lt\{\sum_{\a_1,\a_2,\b_1,\b_2=2}^n
R(u+v+w,\l-w\hat{\imath})^{k\,\,\,\,\b_2}_{\a_2\,\b_1}
R(u-v,\l+w\hat{\jmath})^{\b_1\,\a_1}_{j\,\,\,\,i}\rt.\no\\
&&\qquad\qquad\qquad\qquad\qquad\qquad\qquad\quad
\times\lt.\B_{\b_2}(\l+w(\hat{k}+\hat{\b}_2-\hat{\imath}-\hat{1})|v)
\D^{\a_2}_{\a_1}(\l+w(\hat{\jmath}-\hat{1})|u)\rt\}\no\\
&&\qquad\quad-\frac{\s(w)\s(2u+2w)}{\s(u-v)\s(2u+w)}\lt\{
\sum_{\a,\b=2}^n\frac{\s(u-v+\l_{1\a}-w)}{\s(\l_{1\a}-w)}
\,R(2u+w,\l-w\hat{\imath})^{k\,\b}_{\a\,i}\rt.
\no\\
&&\qquad\qquad\qquad\qquad\qquad\qquad\qquad\qquad\times\lt.
\B_{\b}(\l+w(\hat{k}+\hat{\b}-\hat{\imath}-\hat{1})|u)
\D^{\a}_{j}(\l+w(\hat{\jmath}-\hat{1})|v)\rt\}\no\\
&&\qquad\quad+\frac{\s(w)\s(2v)\s(2u+2w)}{\s(u+v+w)\s(2v+w)\s(2u+w)}\no\\[2pt]
&&\qquad\qquad\qquad\qquad\times\lt\{
\sum_{\a=2}^n\frac{\s(u+v+\l_{1j})}{\s(\l_{1j}-w)}
\,R(2u+w,\l-w\hat{\imath})^{k\,\a}_{j\,i}\rt.
\no\\
&&\qquad\qquad\qquad\qquad\qquad\qquad\times\lt.
\B_{\a}(\l+w(\hat{k}+\hat{\a}-\hat{\imath}-\hat{1})|u)
\A(\l+w(\hat{j}-\hat{1})|v)\rt\},\label{Rel-2}\\[6pt]
&&\B_i(\l+w(\hat{\imath}-\hat{1})|u)
\B_j(\l+w(\hat{\imath}+\hat{\jmath}-2\hat{1})|v)\no\\
&&\qquad=\sum_{\a,\b=2}^nR(u-v,\l-2w\hat{1})^{\b\,\a}_{j\,i}
\,\B_{\b}(\l+w(\hat{\b}-\hat{1})|v)
\,\B_{\a}(\l+w(\hat{\a}+\hat{\b}-2\hat{1})|u).\label{Rel-3}
\end{eqnarray}

In order to apply the algebraic Bethe ansatz method, in addition
to the relevant commutation relations (\ref{Rel-1})-(\ref{Rel-3}),
one needs to construct a pseudo-vacuum state (also called
reference state) which is the common eigenstate of the operators
$\A$, $\D^i_i$ and is annihilated by the operators $\C_i$. In
contrast to the trigonometric and rational cases with {\it
diagonal\/} $K^{\pm}(u)$ \cite{Skl88,Dev94}, the usual
highest-weight state
\begin{eqnarray}
 \lt(\begin{array}{l}1\\0\\\vdots\end{array}\rt)\otimes\cdots\otimes
 \lt(\begin{array}{l}1\\0\\\vdots\end{array}\rt),\no
\end{eqnarray}
is no
longer the pseudo-vacuum state. However, after the face-vertex
transformations (\ref{K-F-1}) and (\ref{K-F-2}), the face type
K-matrices $\K(\l|u)$ and $\tilde{\K}(\l|u)$ {\it
simultaneously\/} become diagonal. This suggests that one can
translate the $\Zb_n$ Belavin model  with  non-diagonal K-matrices
into the corresponding SOS model with {\it diagonal} K-matrices
$\K(\l|u)$ and $\tilde{\K}(\l|u)$ given by
(\ref{K-F-1})-(\ref{K-F-2}). Then one can construct the
pseudo-vacuum in the ``face language" and use the algebraic Bethe
ansatz method  to diagonalize the transfer matrix \cite{Yan04}.

Let us introduce the corresponding pseudo-vacuum state
$|\O\rangle$
\begin{eqnarray}
 |\O\rangle=\phi_{\l-(N-1)w\hat{1},\l-Nw\hat{1}}(-z_1)\otimes
 \phi_{\l-(N-2)w\hat{1},\l-(N-1)w\hat{1}}(-z_{2})\cdots\otimes
 \phi_{\l,\l-w\hat{1}}(-z_N).\label{Vac}
\end{eqnarray}
This state  depends
only on the boundary parameters $\{\l_i\}$ and the inhomogeneous
parameters $\{z_j\}$,  not on the boundary parameter $\xi$. We
find that the pseudo-vacuum state given by (\ref{Vac}) satisfies
the following equations, as required,
\begin{eqnarray}
\A(\l-Nw\hat{1}|u)|\O\rangle&=&k(\l|u;\xi)_1
|\O\rangle,\label{A}\\
\D^i_j(\l-Nw\hat{1}|u)|\O\rangle&=&\d^i_j
f^{(1)}(u)\,k(\l|u+\frac{w}{2};\xi-\frac{w}{2})_j\no\\
&&\hspace{-1mm}\times\!\lt\{\prod_{k=1}^N
\frac{\s(u+z_k)\s(u-z_k)}{\s(u+z_k+w)\s(u-z_k+w)}\!\rt\}\!
|\O\rangle,~i,j=2,\cdots,n,\label{D}\\
\C_i(\l-Nw\hat{1}|u)|\O\rangle&=&0,\qquad i=2,\cdots,n,\\
\B_i(\l-Nw\hat{1}|u)|\O\rangle&\neq& 0,\qquad i=2,\cdots,n. \end{eqnarray} Here
$f^{(1)}(u)$ is given by  \begin{eqnarray} f^{(1)}(u)=
\frac{\s(2u)\s(\l_1+u+w+\xi)}{\s(2u+w)\s(\l_1+u+\xi)}.
\end{eqnarray}

In order to apply the algebraic Bethe ansatz method to diagonalize
the transfer matrix, we need to assume   $N=n\times l$ with $l$
being a positive integer \cite{Yan04}. For convenience, let us
introduce a set of integers: \begin{eqnarray} N_i=(n-i)\times
l,~~i=0,1,\cdots,n-1,\label{Integer} \end{eqnarray} and $\frac{n(n-1)}{2}l$
complex parameters
$\{v^{(i)}_k|~k=1,2,\cdots,N_{i+1},~i=0,1,\cdots,n-2\}$. As in the
usual nested Bethe ansatz method
\cite{Bab82,Sch83,Dev94,Hou03,Yan04}, the parameters
$\{v^{(i)}_k\}$ will be used to specify the eigenvectors of the
corresponding reduced transfer matrices.  They will be constrained
later by the Bethe ansatz equations. For convenience, we adopt the
following convention:
\begin{eqnarray}
 v_k=v^{(0)}_k,~k=1,2,\cdots, N_1.
\end{eqnarray}
We will seek the common eigenvectors (i.e. the so-called Bethe
states) of the transfer matrix in the form
\begin{eqnarray}
|v_1,\cdots,v_{N_1}\rangle &=&\sum_{i_1,\cdots,i_{N_1}=2}^n
F^{i_1,i_2,\cdots,i_{N_1}}\,
\B_{i_1}(\l+w(\hat{\imath}_1-\hat{1})|v_{1})\,\B_{i_2}
(\l+w(\hat{\imath}_1+\hat{\imath}_2-2\hat{1})|v_2)\no\\
&&\qquad\qquad\qquad\times\cdots\B_{i_{N_1}}
(\l+w\sum_{k=1}^{N_1}\hat{\imath}_k-wN_1\hat{1}|v_{N_1})
|\O\rangle.\label{Eigenstate}
\end{eqnarray}
The indices in the above
equation should satisfy the following condition: the number of
$i_k=j$, denoted by $\#(j)$, is $l$, i.e.
\begin{eqnarray}
  \#(j)=l, ~~~j=2, \cdots, n.\label{Restriction}
\end{eqnarray}

With the help of (\ref{De1}), (\ref{Def-AB}) and (\ref{Def-D}) we
rewrite the transfer matrix (\ref{trans}) in terms of the
operators $\A$ and $\D^i_i$
\begin{eqnarray}
 \t(u)&=&\sum_{\nu=1}^n\tilde{k}(\l|u)_{\nu}\,\T(\l|u)^{\nu}_{\nu}
 \no\\
 &=&\tilde{k}(\l|u)_1\,\A(\l|u)+\sum_{i=2}^n\tilde{k}(\l|u)_i
 \,\T(\l|u)^i_i \no\\
 &=&\tilde{k}(\l|u)_1\,\A(\l|u)+ \sum_{i=2}^n\tilde{k}(\l|u)_i
 \,R(2u,\l+w\hat{1})^{i1}_{1i}\,\A(\l|u)\no\\
 &&\phantom{\tilde{k}(\l|u)_1\A(\l|u)}  + \sum_{i=2}^n\tilde{k}(\l|u)_i
 \left(\vTm\T(\l|u)^i_i- R(2u,\l+w\hat{1})^{i1}_{1i}\A(\l|u)\right)\no\\
 &=& \sum_{i=1}^n\tilde{k}(\l|u)_i
 R(2u,\l+w\hat{1})^{i1}_{1i}\,\A(\l|u)\no\\
 &&~~~~~~~~+ \sum_{i=2}^n\tilde{k}^{(1)}(\l|u+\frac{w}{2})_i
 ~\frac{\s(\l_{i1}-w)}{\s(\l_{i1})}\left(\vTm\T(\l|u)^i_i-
 R(2u,\l+w\hat{1})^{i1}_{1i}\A(\l|u)\right)\no\\
 &=&\a^{(1)}(u)\,\A(\l|u)+\sum_{i=2}^n
 \tilde{k}^{(1)}(\l|u+\frac{w}{2})_i\,\D(\l|u)^i_i.
 \label{trans1}
\end{eqnarray}
Here we have used (\ref{Def-D}) and introduced
the function $\a^{(1)}(u)$,
\begin{eqnarray}
 \a^{(1)}(u)=\sum_{i=1}^n\tilde{k}(\l|u)_i\,
 R(2u,\l+w\hat{1})^{i1}_{1i},\label{function-a}
\end{eqnarray}
and the reduced
 K-matrix $\tilde{\K}^{(1)}(\l|u)$ with the elements given by
\begin{eqnarray}
\tilde{\K}^{(1)}(\l|u)^j_i&=&
 \d^j_i\,\tilde{k}^{(1)}(\l|u)_i,\qquad\qquad\qquad
 i,j=2,\cdots,n,\label{Reduced-K1}\\
\tilde{k}^{(1)}(\l|u)_i&=& \lt\{\prod_{k\neq
 i,k=2}^n\frac{\s(\l_{ik}-w)}{\s(\l_{ik})}\rt\}
 \frac{\s(\l_i+\bar{\xi}+u+
 \frac{(n-1)w}{2})}{\s(\l_i+\bar{\xi}-u-\frac{(n-1)w}{2})}.
 \label{Reduced-K2}
\end{eqnarray}
To carry out the nested Bethe ansatz
process \cite{Bab82,Sch83,Dev94,Hou03,Yan04} for  the $\Zb_n$
Belavin model with the generic open boundary conditions, one needs
to introduce a set of reduced K-matrices $\tilde{\K}^{(m)}(\l|u)$
\cite{Yan04} which include the original one
$\tilde{\K}(\l|u)=\tilde{\K}^{(0)}(\l|u)$ and the ones in
(\ref{Reduced-K1}) and (\ref{Reduced-K2}):
\begin{eqnarray}
 \tilde{\K}^{(m)}(\l|u)^j_i&=&
 \d^j_i\,\tilde{k}^{(m)}(\l|u)_i,\quad i,j=m+1,\cdots,n,
 \quad m=0,\cdots,n-1,
 \label{Reduced-K3}\\
 \tilde{k}^{(m)}(\l|u)_i&=& \lt\{\prod_{k\neq
 i,k=m+1}^n\frac{\s(\l_{ik}-w)}{\s(\l_{ik})}\rt\}
 \frac{\s(\l_i+\bar{\xi}+u+
 \frac{(n-m)w}{2})}{\s(\l_i+\bar{\xi}-u-\frac{(n-m)w}{2})}.
 \label{Reduced-K4}
\end{eqnarray}
Moreover we introduce a set of functions
$\{\a^{(m)}(u)|m=1,\cdots,n-1\}$ (including the one in
(\ref{function-a})) related to the reduced K-matrices
$\tilde{\K}^{(m)}(\l|u)$
\begin{eqnarray}
 \a^{(m)}(u)=\sum_{i=m}^{n}R(2u,\l+w\hat{m})^{im}_{mi}\,
 \tilde{k}^{(m-1)}(\l|u)_i,\quad m=1,\cdots,n.\label{function-a-1}
\end{eqnarray}

Carrying out the nested Bethe ansatz,  we find \cite{Yan04} that
with the coefficients $F^{i_1,i_2,\cdots,i_{N_1}}$ in
(\ref{Eigenstate}) properly chosen,  the Bethe state
$|v_1,\cdots,v_{N_1}\rangle $ is the eigenstate of the transfer
matrix (\ref{trans}),
\begin{eqnarray}
 \t(u)|v_1,\cdots,v_{N_1}\rangle=\L(u;\xi,\{v_k\})
 |v_1,\cdots,v_{N_1}\rangle,\end{eqnarray} with eigenvalues given by
 \begin{eqnarray}
 \L(u;\xi,\{v_k\})&=&\a^{(1)}(u)\frac{\s(\l_1+\xi-u)}
 {\s(\l_1+\xi+u)}\prod_{k=1}^{N_1}\frac{\s(u+v_k)\s(u-v_k-w)}
 {\s(u+v_k+w)\s(u-v_k)}\no\\
 &&+\frac{\s(2u)\s(\l_1+u+w+\xi)} {\s(2u+w)\s(\l_1+u+\xi)}
 \prod_{k=1}^{N_1}\frac{\s(u-v_k+w)\s(u+v_k+2w)}
 {\s(u-v_k)\s(u+v_k+w)}\no\\
 &&\ \ \times \prod_{k=1}^{N_0}\frac{\s(u+z_k)\s(u-z_k)}
 {\s(u+z_k+w)\s(u-z_k+w)}
 \L^{(1)}(u+\frac{w}{2};\xi-\frac{w}{2},\{v^{(1)}\}).\label{Eigenvalue1}
\end{eqnarray}
The eigenvalues $\{\L^{(i)}(u;\xi,\{v^{(i)}_{k}\})\}$ (with the original one
$\L(u;\xi,\{v_{k}\})=\L^{(0)}(u;\xi,\{v^{(0)}_{k}\})$) of the
reduced transfer matrices are given by the following recurrence
relations
\begin{eqnarray}
 \L^{(i)}(u;\xi^{(i)},\{v^{(i)}_k\})&=&\a^{(i+1)}(u)
 \frac{\s(\l_{i+1}+\xi^{(i)}-u)} {\s(\l_{i+1}+\xi^{(i)}+u)}
 \prod_{k=1}^{N_{i+1}}\frac{\s(u+v^{(i)}_k)\s(u-v^{(i)}_k-w)}
 {\s(u+v^{(i)}_k+w)\s(u-v^{(i)}_k)}\no\\
 &&+\frac{\s(2u)\s(\l_{i+1}+u+w+\xi^{(i)})}
 {\s(2u+w)\s(\l_{i+1}+u+\xi^{(i)})}
 \prod_{k=1}^{N_{i+1}}\frac{\s(u-v^{(i)}_k+w)\s(u+v^{(i)}_k+2w)}
 {\s(u-v^{(i)}_k)\s(u+v^{(i)}_k+w)}\no\\
 &&\,\times
 \prod_{k=1}^{N_i}\frac{\s(u+z^{(i)}_k)\s(u-z^{(i)}_k)}
 {\s(u+z^{(i)}_k+w)\s(u-z^{(i)}_k+w)}
 \L^{(i+1)}(u+\frac{w}{2};\xi^{(i)}-\!\frac{w}{2},\{v^{(i+1)}\}),\no\\
 &&~~~~~~~~~~~~i=1,\cdots,n-2, \label{Eigenvalue2}\\
 \L^{(n-1)}(u;\xi^{(n-1)})&=&\frac{\s(\l_n+\bar{\xi}+u+\frac{w}{2})
 \s(\l_n+\xi^{(n-1)}-u)} {\s(\l_n+\bar{\xi}-u-\frac{w}{2})
 \s(\l_n+\xi^{(n-1)}+u)}.\label{Eigenvalue3}
\end{eqnarray}
The reduced
boundary parameters $\{\xi^{(i)}\}$ and inhomogeneous parameters
$\{z^{(i)}_k\}$ are given by
\begin{eqnarray}
 \xi^{(i+1)}=\xi^{(i)}-\frac{w}{2},
 \qquad z^{(i+1)}_k=v^{(i)}_k+\frac{w}{2},
 \qquad i=0,\cdots,n-2.\label{Parameters}
\end{eqnarray}
Here we have adopted the
convention: $\xi=\xi^{(0)}$, $z^{(0)}_k=z_k$. The complex
parameters  $\{v^{(i)}_k\}$ satisfy the following Bethe ansatz
equations:
\begin{eqnarray}
 &&\hspace{-10pt}\a^{(1)}(v_s)\frac{\s(\l_1+\xi-v_s)\s(2v_s+w)}
 {\s(\l_1+\xi+v_s+w)\s(2v_s+2w)} \prod_{k\ne s,k=1}^{N_1}
 \frac{\s(v_s+v_k)\s(v_s-v_k-w)} {\s(v_s+v_k+2w)\s(v_s-v_k+w)}
 \no\\
 &&\qquad\qquad = \prod_{k=1}^{N} \frac{\s(v_s+z_k)\s(v_s-z_k)}
 {\s(v_s+z_k+w)\s(v_s-z_k+w)}
 \,\L^{(1)}(v_s+\frac{w}{2};\xi-\frac{w}{2},\{v^{(1)}_k\}),\qquad\qquad
 \label{BA1}\\
 &&\hspace{-10pt}\a^{(i+1)}(v^{(i)}_s)\frac{\s(\l_{i+1}+\xi^{(i)}-v^{(i)}_s)
 \s(2v^{(i)}_s+w)}
 {\s(\l_{i+1}+\xi^{(i)}+v^{(i)}_s+w)\s(2v^{(i)}_s+2w)}\!\!
 \prod_{k\ne s,k=1}^{N_{i+1}}
 \frac{\s(v^{(i)}_s+v^{(i)}_k)\s(v^{(i)}_s-v^{(i)}_k-w)}
 {\s(v^{(i)}_s+v^{(i)}_k+2w)\s(v^{(i)}_s-v^{(i)}_k+w)}\no\\
 &&\qquad\qquad = \prod_{k=1}^{N_i}
 \frac{\s(v^{(i)}_s+z^{(i)}_k)\s(v^{(i)}_s-z^{(i)}_k)}
 {\s(v^{(i)}_s+z^{(i)}_k+w)\s(v^{(i)}_s-z^{(i)}_k+w)}
 \,\L^{(i+1)}(v^{(i)}_s+\frac{w}{2};\xi^{(i)}-\frac{w}{2},\{v^{(i+1)}_k\})
 ,\no\\
 &&\qquad\qquad\qquad\qquad\qquad\qquad i=1,\cdots,n-2.
 \label{BA2}
\end{eqnarray}

\subsection{Eigenvalues and associated Bethe ansatz equations}
The relation (\ref{Eq-1}) between $\{H_j\}$ and $\{\t(z_j)\}$ and
the fact that the first term on the r.h.s. of (\ref{trans-2}) is a
c-number enable us to extract the eigenstates  of the elliptic
Gaudin operators and the corresponding eigenvalues from the
results obtained in the previous subsection.

Introduce the functions $\{\b^{(i)}(u,w)\}$
\begin{eqnarray}
 \b^{(i+1)}(u,w)\equiv
 \b^{(i+1)}(u)=\a^{(i+1)}(u)\frac{\s(\l_{i+1}+\xi-u-\frac{i}{2}w)}
 {\s(\l_{i+1}+\xi+u+w-\frac{i}{2}w)},~~i=0,\cdots,
 n-2.\label{function-b}
\end{eqnarray}
Then by (\ref{Eigenvalue2}),
(\ref{Eigenvalue3}) and (\ref{Parameters}), the Bethe ansatz
equations (\ref{BA1}) and (\ref{BA2}) become, respectively,
\begin{eqnarray}
 &&\hspace{-24pt} \b^{(i+1)}(v^{(i)}_s)\frac{\s(2v^{(i)}_s+w)}{\s(2v^{(i)}_s+2w)}
 \prod_{k\neq s,k=1}^{N_{i+1}}\frac{\s(v^{(i)}_s+v^{(i)}_k)
 \s(v^{(i)}_s-v^{(i)}_k-w)}{\s(v^{(i)}_s+v^{(i)}_k+2w)
 \s(v^{(i)}_s-v^{(i)}_k+w)}\no\\
 && ~~~~=\b^{(i+2)}(v^{(i)}_s+\frac{w}{2})
 \prod_{k=1}^{N_{i}}\frac{\s(v^{(i)}_s+v^{(i-1)}_k+\frac{w}{2})
 \s(v^{(i)}_s-v^{(i-1)}_k-\frac{w}{2})}{\s(v^{(i)}_s+v^{(i-1)}_k+\frac{3w}{2})
 \s(v^{(i)}_s-v^{(i-1)}_k+\frac{w}{2})}\no\\
 &&\qquad\qquad\qquad\qquad\
 \times\prod_{k=1}^{N_{i+2}}\frac{\s(v^{(i)}_s+v^{(i+1)}_k+\frac{w}{2})
 \s(v^{(i)}_s-v^{(i+1)}_k-\frac{w}{2})}{\s(v^{(i)}_s+v^{(i+1)}_k+\frac{3w}{2})
 \s(v^{(i)}_s-v^{(i+1)}_k+\frac{w}{2})},\no\\
&&\hspace{22mm}i=0,\cdots,n-3,\label{BA1-1}\\
 &&
 \hspace{-24pt}
 \b^{(n-1)}(v^{(n-2)}_s)\frac{\s(2v^{(n-2)}_s+w)}{\s(2v^{(n-2)}_s+2w)}
 \prod_{k\neq s,k=1}^{N_{n-1}}\frac{\s(v^{(n-2)}_s+v^{(n-2)}_k)
 \s(v^{(n-2)}_s-v^{(n-2)}_k-w)}{\s(v^{(n-2)}_s+v^{(n-2)}_k+2w)
 \s(v^{(n-2)}_s-v^{(n-2)}_k+w)}\no\\
 &&~~~~=\frac{\s(\l_n+\xi+v^{(n-2)}_s+w)\s(\l_n+\xi^{(n-1)}-v^{(n-2)}_s-\frac{w}{2})}
 {\s(\l_n+\xi-v^{(n-2)}_s-w)\s(\l_n+\xi^{(n-1)}+v^{(n-2)}_s+\frac{w}{2})}\no\\
&&\qquad\qquad\qquad\qquad
\times\prod_{k=1}^{N_{n-2}}\frac{\s(v^{(n-2)}_s+v^{(n-3)}_k+\frac{w}{2})
\s(v^{(n-2)}_s-v^{(n-3)}_k-\frac{w}{2})}{\s(v^{(n-2)}_s+v^{(n-3)}_k+\frac{3w}{2})
 \s(v^{(n-2)}_s-v^{(n-3)}_k+\frac{w}{2})}.\label{BA1-2}
\end{eqnarray}
Here
we have used the convention: $v^{(-1)}_k=z_k,\, k=1,\cdots,N$. The
quasi-classical property (\ref{quasi1}) of  $R(u,m)$,
(\ref{function-a-1}) and (\ref{function-b}) lead to the following
relations
\begin{eqnarray}
 \b^{(i+1)}(u,0)=1,\quad \frac{\partial}{\partial
 u}\b^{(i+1)}(u,0)=0,\quad i=0,\cdots,n-2.\label{functionb-1}
\end{eqnarray}
Then,
one may  introduce functions $\{\g^{(i+1)}(u)\}$ associated with
$\{\b^{(i+1)}(u,w)\}$
\begin{eqnarray}
 \g^{(i+1)}(u)=\frac{\partial}{\partial
 w}\b^{(i+1)}(u,w)|_{w=0},\quad i=0,\cdots,n-2.\label{fuctiong}
\end{eqnarray}
Using (\ref{functionb-1}), we can extract the eigenvalues $h_j$
(resp. the corresponding Bethe ansatz equations) of the Gaudin
operators $H_j$ (\ref{Ham}) from  the expansion around $w=0$ for
the first order of $w$ of the eigenvalues (\ref{Eigenvalue1}) of
the transfer matrix $\t(u=z_j)$ (resp. the Bethe ansatz equations
(\ref{BA1-1}) and (\ref{BA1-2}) ). Finally, the eigenvalues of the
$\Zb_n$ elliptic Gaudin operators are
\begin{eqnarray}
h_j=\g^{(1)}(z_j)+\zeta(\l_1+\xi+z_j)-\sum_{k=1}^{N_1}
\lt\{\zeta(z_j+x_k)+\zeta(z_j-x_k)\rt\},\label{Eig-1}\end{eqnarray} where
$\zeta$-function is defined in (\ref{Z-function}). The
$\frac{n(n-1)l}{2}$ parameters
$\{x^{(i)}_k|~k=1,2,\cdots,N_{i+1},~i=0,1,\cdots,n-2\}$ (including
$x_k$ as $x_k=x^{(0)}_k,\,k=1,\cdots, N_1$) are determined by the
following associated Bethe ansatz equations
\begin{eqnarray}
 &&\g^{(i+1)}(x^{(i)}_s)-\zeta(2x^{(i)}_s)-2\sum_{k\neq
 s,k=1}^{N_{i+1}}\lt\{\zeta(x^{(i)}_s+x^{(i)}_k)+
 \zeta(x^{(i)}_s-x^{(i)}_k)\rt\}\no\\
 &&\qquad\qquad\quad =\g^{(i+2)}(x^{(i)}_k)-\sum_{k=1}^{N_{i}}
 \lt\{\zeta(x^{(i)}_s+x^{(i-1)}_k)+
 \zeta(x^{(i)}_s-x^{(i-1)}_k)\rt\}\no\\
 &&\qquad\qquad\qquad\qquad\qquad\ \  -\sum_{k=1}^{N_{i+2}}
 \lt\{\zeta(x^{(i)}_s+x^{(i+1)}_k)+
 \zeta(x^{(i)}_s-x^{(i+1)}_k)\rt\},\no\\
&&\hspace{27mm}i=0,\cdots,n-3,\label{BAE-1}\\
 &&\g^{(n-1)}(x^{(n-2)}_s)-\zeta(2x^{(n-2)}_s)-2\sum_{k\neq
 s,k=1}^{N_{n-1}}\lt\{\zeta(x^{(n-2)}_s+x^{(n-2)}_k)+
 \zeta(x^{(n-2)}_s-x^{(n-2)}_k)\rt\}\no\\
 &&\qquad\qquad\quad =\frac{n}{2}\zeta(\l_n+\xi+x^{(n-2)}_s)+
 \frac{2-n}{2}\zeta(\l_n+\xi-x^{(n-2)}_s)\no\\
 &&\qquad\qquad\qquad\qquad\qquad\quad\quad -\sum_{k=1}^{N_{n-2}}
 \lt\{\zeta(x^{(n-2)}_s+x^{(n-3)}_k)+
 \zeta(x^{(n-2)}_s-x^{(n-3)}_k)\rt\}.\label{BAE-2}
\end{eqnarray}
Here we
have used the convention: $x^{(-1)}_k=z_k,\,k=1,\cdots,N$ in
(\ref{BAE-1}).

\section{Conclusions} \label{Con}
\setcounter{equation}{0}

We have studied the $\Zb_n$ elliptic  Gaudin model with generic
boundaries specified by the non-diagonal K-matrices $K^{\pm}(u)$,
(\ref{K-matrix}) and (\ref{K-matrix1}). In addition to the
inhomogeneous parameters $\{z_j\}$, the associated Gaudin
operators $\{H_j\}$, (\ref{Ham}), have $n+1$ free parameters
$\{\l_i\}$ and $\xi$. As seen from section 4, although the
``vertex type" K-matrices $K^{\pm}(u)$ (\ref{K-matrix}) and
(\ref{K-matrix1}) are {\it non-diagonal\/}, the compositions,
(\ref{K-F-1}) and (\ref{K-F-2}), lead to the {\it diagonal\/}
``face-type" K-matrices after the face-vertex transformation. This
enables us to successfully construct the corresponding
pseudo-vacuum state $|\O\rangle$ (\ref{Vac}) and apply the
algebraic Bethe ansatz method to diagonalize the transfer matrix
$\t(u)$ of the inhomogeneous $\Zb_n$ Belavin model with generic
open boundaries. Furthermore, we have exactly diagonalized the
generalized Gaudin operators $\{H_j\}$, and derived their
eigenvalues (\ref{Eig-1}) as well as the associated Bethe ansatz
equations (\ref{BAE-1}) and (\ref{BAE-2}). Taking the scaling
limit \cite{Yan01} of our general results for the special $n=2$
case, we recover the results obtained in \cite{Yan041} for the XXZ
Gaudin model with generic open boundaries.

\section*{Acknowledgements}
The financial support from  Australian Research Council through
Discovery-Projects and Linkage-International grants  is gratefully
acknowledged.


\end{document}